\pdfoutput=1
\documentclass[conference]{IEEEtran}
\usepackage{cite}
\usepackage{comment}
\usepackage{algorithmic}
\usepackage{xcolor}
\usepackage{graphicx}
\usepackage{textcomp}
\usepackage{tabularx}
\usepackage{xcolor}
\usepackage{lipsum}
\usepackage{dblfloatfix}
\usepackage{tabularx}
\usepackage{comment}
\usepackage{amsmath}
\usepackage{bm}
\usepackage[singlelinecheck=false,justification=justified]{caption}

\def\BibTeX{{\rm B\kern-.05em{\sc i\kern-.025em b}\kern-.08em
    T\kern-.1667em\lower.7ex\hbox{E}\kern-.125emX}}
\graphicspath{{./figures/}}

\def\BibTeX{{\rm B\kern-.05em{\sc i\kern-.025em b}\kern-.08em
    T\kern-.1667em\lower.7ex\hbox{E}\kern-.125emX}}

\definecolor{mycommentcolor1}{RGB}{255,0,0} 
\definecolor{mycommentcolor2}{RGB}{0,0,255} 

\begin{document}
\title{Resilient Temporal GCN for Smart Grid State Estimation Under Topology Inaccuracies}

\author{\IEEEauthorblockN{Seyed Hamed Haghshenas\IEEEauthorrefmark{1},
Mia Naeini\IEEEauthorrefmark{2}}
\IEEEauthorblockA{Department of Electrical Engineering, University of South Florida, Tampa, Florida\\
Email: \IEEEauthorrefmark{1}seyedhamedhaghshenas@usf.edu,
\IEEEauthorrefmark{2}mnaeini@usf.edu}}
\maketitle

\begin{abstract}
State Estimation is a crucial task in power systems. Graph Neural Networks have demonstrated significant potential in state estimation for power systems by effectively analyzing measurement data and capturing the complex interactions and interrelations among the measurements through the system's graph structure. However, the information about the system's graph structure may be inaccurate due to noise, attack or lack of accurate information about the topology of the system. This paper studies these scenarios under topology uncertainties and evaluates the impact of the topology uncertainties on the performance of a Temporal Graph Convolutional Network (TGCN) for state estimation in power systems. In order to make the model resilient to topology uncertainties, modifications in the TGCN model are proposed to incorporate a knowledge graph, generated based on the measurement data. This knowledge graph supports the assumed uncertain system graph. Two variations of the TGCN architecture are introduced to integrate the knowledge graph, and their performances are evaluated and compared to demonstrate improved resilience against topology uncertainties. The evaluation results indicate that while the two proposed architecture show different performance, they both improve the performance of the TGCN state estimation under topology uncertainties.
\end{abstract}

\begin{IEEEkeywords}
Graph Topology Noise, Smart Grids, State Estimation, Graph Neural Networks.
\end{IEEEkeywords}

\IEEEpeerreviewmaketitle

\section{Introduction}
State Estimation (SE) is a critical function in power systems, essential for effective monitoring, planning, and operations. With the large deployment of measurement devices, such as Phasor Measurement Units (PMUs), vast amount of data is generated that are being utilized through data-driven methods for SE \cite{Data-DrivenSE1, Data-DrivenSE2, Data-DrivenSE3} to complement the conventional model-based SE techniques \cite{SEbook1, SEbook2}. Recently, machine learning (ML) models for SE have gained significant attention. 

Given that measurement data from power systems inherently possess structures due to the underlying physical network of the system and physics of the electricity, Graph Neural Networks (GNNs) have been particularly effective. In GNN models, both the measurement data and the system's graph are essential components for constructing the model. The measurement data provides information on the state of the power system, while the graph represents the system's topology, specifying how different nodes (such as buses, substations, and generators) are interacting. This combination allows GNN models to effectively capture and leverage the complex interactions and dependencies inherent in the power system, leading to more accurate and robust state estimation.

In GNN, both the measurement data and the system's graph are vulnerable to inaccuracies caused by noise, attacks, or system failures. While numerous studies have assessed the impact of noise and cyber attacks, such as False Data Injection Attacks (FDIAs), in the system measurements on SE models \cite{TGCN-Attack}, the effects of uncertainties in the system's graph information due to noise, attacks or failures remain underexplored and inadequately addressed. While topology uncertainties can affect the links, nodes, and connection weights within the graph, this paper primarily focuses on uncertainties impacting the links. Such uncertainties may arise from noise that alters the perceived status of a link to either up or down in the system's topology, or from deliberate attacks on topology information designed to mislead GNN models. 

This paper aims to study the effects of the aforementioned topology uncertainties on a specific Temporal Graph Convolutional Network (TGCN) model for SE in power systems initially presented in \cite{mdjakirtgcn}. Next, in order to make the model resilient to topology uncertainties, modifications in the TGCN model are proposed to incorporate a knowledge graph, generated based on the measurement data. This knowledge graph supports the assumed uncertain system graph. Three methods for defining the knowledge graph are explored in the design of the resilient TGCN. Additionally, two variations of the TGCN architecture are introduced for integrating the knowledge graph in the model. The performance of the proposed resilient TGCN models is evaluated across the two architecture variations and the three knowledge graph designs and compared with the original TGCN model for SE under topology uncertainties. The results demonstrate that while all the studied variations improve the TGCN's performance under topology uncertainties, certain configurations offer greater improvement.

Our main contributions are summarized as follows:
\begin{itemize}
    \item The effects of topology uncertainties in the form of link removal and link addition are evaluated for the TGCN-based SE initially proposed in \cite{TopologyAttack2}.
    \item A resilient TGCN model, namely, Knowledge Graph Infused Model (KGIM) SE, is proposed based on integrating a measurement-based knowledge graph into the original TGCN model. This model augments the uncertain graph with the knowledge graph, which improves message passing within the TGCN, improving the performance under topology uncertainties. 
    \item A variation of the KGIM, called the Parallel Knowledge Graph Infused Model (PKGIM), is designed to integrate state estimation from two parallel TGCN channels: one utilizing the uncertain graph and the other utilizing the knowledge graph. The output features from these channels are combined using methods such as Single Layer Perceptron (SLP), Multi-Layer Perceptron (MLP) Aggregation, and Attention Aggregation.
    \item Three variations of the knowledge graph are evaluated for the KGIM and PKGIM models: 1) a graph based on cosine similarity between measurements, 2) a graph using a Graph Attention Network (GAT) to capture similarities among measurements, and 3) a graph based on Pearson Correlation between measurements.
    \item The performance of the proposed resilient TGCN models is assessed across two architectural variations and three knowledge graph designs. This evaluation is compared with the original TGCN model for SE under topology uncertainties, using a comprehensive set of graph uncertainty scenarios with link additions and link removals, applied to the IEEE 118-bus system.
\end{itemize}

\section{Related Work}\label{relatedworks}
SE is a crucial function that enhances situational awareness, ensuring the efficient and reliable operation of power systems. Conventional SE methods are generally physics-based models that rely heavily on the system model \cite{SEbook1, SEbook2, SEpaper}. Recently, research on data-driven SE \cite{Data-DrivenSE1,Data-DrivenSE2,Data-DrivenSE3} has grown, aiming to complement and enhance the conventional techniques. These approaches often use ML to process the abundance of measurement data from these systems and improve SE performance under system model changes \cite{NN-SE1,AiSmartGrids}, missing or inaccurate information \cite{TopologyAttackDetectionGSP}, and physical and cyber stresses \cite{mdjakirlfd, zhang21}. Among various ML techniques GNNs are gaining more attentions recently due to their abilities in capturing the underlying interactions among the measurement data from the power systems \cite{TGCN-Attack, boyaci22jan,boyaci22jun, boyaci21oct}. While numerous studies have evaluated the effects of noise and cyber attacks, such as FDIA, on power system measurements and SE models \cite{TGCN-Attack,TopologyAttack2,boyaci22jan,boyaci22jun,boyaci21oct,chu19,hasnat21}, research on the impact of uncertainties in the system's graph on GNN models remains limited. For instance, recent works in \cite{Zugner, TopologyAttack1, liu, he} have assessed the vulnerability of general GNN models to topology information attack and noise; showing that adding or removing links can significantly degrade GNN performance in tasks such as node classification, graph representation learning and link prediction in applications including social network\cite{TKG}, citation network\cite{Zugner,he,NodeFormer}, water systems\cite{GNNAnomalyDetection} and energy management systems\cite{rahman}. Some of such studies are focused more on topology attack detection\cite{TopologyAttackDetection1,GNNAnomalyDetection} and some focused on developing defense against topology attacks\cite{TopologyAttack1,liu,GNNGuard} and mitigating the impact of topology noise in GNNs\cite{he,luo,dai}.

In power systems, the works presented in \cite{rahman,TopologyAttack2,TopologyAttackDetection2,kim,jolfaei,moshtagh} studied the effects of topology noise and attacks on various functions, such as SE and cyber stress detection. For instance, the work presented in \cite{rahman} examines the impact of topology poisoning attacks on the economic operation of smart grids, particularly optimal power flow. The impact of topology noise or attack on SE was investigated in \cite{TopologyAttack2}. In this study, a temporal GNN framework is used to perform the SE task in the presence of topology noise, such as missing or extra links in the topology. According to \cite{TopologyAttack2} topology noise or attack in power systems can be modeled as FDI attacks on topology information. For instance, in \cite{TopologyAttackDetectionGSP}, a Graph Signal Processing (GSP) technique was used to detect and locate topology attack in power systems. This GSP-based technique relies on local and global smoothness features of the graph signals, which enables detecting and locating FDI attacks on the graph's topology.

Other studies on graph topology attacks, such as \cite{kim}, examine how man-in-the-middle attacks manipulate meter and network switch data to mislead the control center. To defend against such attacks, \cite{kim} proposes securing a subset of meters that are critical to the grid's observability. This approach is known as ``cover-up-strategy" which involves protecting specific line flow meters and bus injection meters to ensure that attacks will be detected. The study presented in \cite{jolfaei} explores an advanced persistent threat scenario, where attackers gradually alter the power grid's topology to disrupt SE. It proposes a detection method based on the expected energy of normalized residues and a paired t-test. This method is designed to detect the cumulative effect of multiple small perturbations, identifying anomalies over several SE cycles.

\textcolor{black}{The authors in \cite{moshtagh} introduce a time-synchronized GNN-based SE framework that exploits a GCN to aggregate information from neighboring nodes and a GAT to focus on important neighboring information by adjusting their weight.The GAT layer helps the GNN model adapt to topology changes by prioritizing the most relevant neighboring nodes, thus making the model more resilient to changes in the system topology.}

Creating and utilizing a knowledge graph based on system measurements is another approach that has been used in topology noise and abnormality detection in GNNs \cite{TopologyIdentification} and fault diagnosis \cite{BigData-KG} in power systems. This approach involves developing a structured representation of information as a network of entities and their relationships. For example, in \cite{TopologyIdentification}, knowledge graphs derived from system measurements serve as the sole input to a GNN for distinguishing between the original and noisy grid topologies. In another study \cite{BigData-KG}, knowledge graphs are constructed using extensive data from power system measurements to identify and diagnose faults within electrical power systems. By leveraging these information-based graphs, this work aims to enhance the accuracy and effectiveness of fault detection and diagnosis processes.

The most similar work to the current paper is presented in \cite{TopologyAttackDetection2}, which proposes a robust SE method for distribution networks with inaccurate topology information. In the latter model, a knowledge graph derived using an aggregated \textit{k}-nearest neighbor approach and the original inaccurate graphs are input into an adaptive multi-channel GAT model to fuse graph embeddings. While our proposed model also incorporates a knowledge graph into the SE framework, it differs in its GNN-based models, including the architecture, fusion process, and knowledge graph derivation. Furthermore, \cite{TopologyAttackDetection2} evaluates the proposed model under only three topology inaccuracy scenarios on the IEEE 33-node system, whereas our study examines a broader range of graph uncertainty scenarios, including random link addition and removal, on the IEEE 118-bus system, presenting a more comprehensive analysis of robustness and vulnerabilities against topology uncertainties.

\section{Problem Formulation}
\subsection{Power Systems as Graphs}
In this paper, the physical structure of the power system is represented as a graph $\mathcal G=(\mathcal{V},\mathcal{E})$, where $\mathcal{V}$ is the set of vertices representing the buses (total $N$ buses), and $\mathcal{E}=\{e_{ij}: (i,j) \in \mathcal{V} \times \mathcal{V}\}$ is the set of edges (total $M$ edges) representing the transmission lines of the power system. The connectivity in the graph can be captured in the adjacency matrix, $\bm{A} = [a_{ij}]$, where $a_{ij}=1$ for $e_{ij}\in \mathcal{E}$ and zero, otherwise. A graph signal at time instance $t$ is denoted by $S(n,t)$ and is defined by assigning values corresponding to the attributes of interest to each node, $n \in \mathcal{V}$ in the graph, thereby establishing a signal over the domain $\mathcal{V}$. Various attributes can be associated with each bus, for instance, the bus voltage magnitude and phase angle, real and reactive power injections, injected bus current magnitude and angle, and frequency. This work considers the phase angle at the buses as a time-varying graph signal. It is assumed that the measurements are available at all the buses of the system through PMUs.

\subsection{Topology Inaccuracies}
As discussed earlier, in graph-aware analysis of data from power systems, the system’s graph information is a crucial component of the model; however, it is vulnerable to inaccuracies caused by noise, attacks, or missing information. Specifically, noise in topology information can arise from limited observability in the system due to missing, failed, or malfunctioning sensing and monitoring devices, as well as from failed or noisy communication links that cannot accurately capture the true state of system components (e.g., unrecorded transmission line trips). Such inaccuracies are random in nature.
On the other hand, attacks on topology information involve the deliberate injection of incorrect data about the components or their connections within the system's graph that can mislead GNN and GSP-based models.

In this work, the notation $\mathcal{G}' = (\mathcal{V}', \mathcal{E}')$ is used to represent the inaccurate topology, as opposed to the true system topology $\mathcal{G} = (\mathcal{V}, \mathcal{E})$ defined earlier. While both node and edge information are vulnerable to inaccuracies due to noise or attack, this work focuses specifically on inaccuracies in the graph's links. It is assumed that $\mathcal{E}'$ differs from $\mathcal{E}$ by the presence of a few extra or missing links. Furthermore, it is assumed that there are either extra links or missing links, but not both simultaneously. Exploring the impact of having both missing and extra links in the topology could be a potential direction for future research. Here, $e^{E}_{ij}$ denotes an extra edge mistakenly present between nodes $i$ and $j$, and $e^{M}_{ij}$ denotes a missing edge that originally existed between nodes $i$ and $j$ in the graph.
In this model with inaccurate topology information, the adjacency matrix changes to $\bm{A'}$, which can disrupt information sharing, message passing, and graph convolutional functions in GNN models. This work examines the effects of such inaccuracies, including missing and extra links in the graph model, on SE performance in power systems using a TGCN model. 

\subsection{A Review of TGCN Model for State Estimation}\label{tgcn}
This study builds upon the TGCN framework, first introduced in~\cite{mdjakirtgcn}. In this section, a brief overview of the framework is presented to set the stage for the following sections, where the effects of uncertain topology on SE using TGCN is studied. In this study, phase angle $x_n$, defines the state of node $n$ and the system state for all the nodes at time $t$ is denoted by {$\bm{X}_{t}:=[{\bm{x}}_{n,t}]^T$}, which is a vector of size $N$. The model performs one-step-ahead state prediction by capturing both the spatial and temporal interactions in the data through message passing within neighborhoods and a Gated Recurrent Unit (GRU), respectively. 

As such, the TGCN framework comprises two layers. The first layer is a graph convolution layer that utilizes a message passing framework to effectively capture the system's structure and interactions among its components. The graph convolution layer can be formulated as $\mathcal{H}^{l+1} = \sigma(\tilde{\bm D}^{-\frac{1}{2}}\tilde{\bm A}\tilde{\bm D}^{-\frac{1}{2}}\mathcal{H}^l\mathcal{\bm W}_l)$, where $\tilde{\bm A}:=\bm{A}+\bm I_N$, with $\bm A$ representing the adjacency matrix and $\bm I_N$ being the identity matrix of size $N$. Additionally, $\tilde{\bm D}:=I_N\sum_{j}{\tilde{\bm A}_{i,j}}$ serves as the degree matrix. Here, $\sigma(\cdot)$ denotes the sigmoid activation function, and $\mathcal{H}^l$ represents the output of layer $l$ with weights $\mathcal{\bm W}_l$. \textcolor{black}{In graph convolution layer, the number of layers, $l$, is set to two 
with the output layer corresponding to $l=2$.}
The second layer of the TGCN framework is a GRU layer, which is responsible for capturing temporal dependencies within the time-series data. \textcolor{black}{Stacking the GRU model with the GCN forms a temporal GCN, initially introduced in \cite{GCN-GRU}. The TGCN procedure can be outlined as follows.}
\begin{equation}
\begin{aligned}
    &F(\bm X_t,\bm A) = \sigma(\hat{\bm A}\text{ReLU}(\hat{\bm A}\bm X_t\mathcal{\bm W}_0)\mathcal{\bm W}_1) \\
    &u_t = \sigma_{1}(\mathcal{\bm W}_u[F(\bm X_t,\bm A), h_{t-1}]+ b_u) \\
    &r_t = \sigma_{1}(\mathcal{\bm W}_r[F(\bm X_t,\bm A), h_{t-1}]+ b_r) \\
    &z_t = \sigma_{2}(\mathcal{\bm W}_z[F(\bm X_t,\bm A), (r_t \odot h_{t-1})]+ b_z) \\
    &h_t = u_t \odot h_{t-1} + (1-u_t) \odot z_t \\
\end{aligned}
\end{equation}

where $\hat{\bm A} := \tilde{\bm D}^{-\frac{1}{2}}\tilde{\bm A}\tilde{\bm D}^{-\frac{1}{2}}$ and $\mathcal{\bm W}_0 \in {R}^{\beta\times\delta}$ and $\mathcal{\bm W}_1 \in {R}^{\delta\times\tau}$ are the model weights with $\beta$, $\delta$, and $\tau$ representing the batch size, hidden units, and prediction length, respectively.
\textcolor{black}{In this model, $r_t$ represents the reset gate, controlling how much past information should be discarded, while $u_t$ is the update gate, determining the amount of past information to carry forward. Additionally, $z_t$ serves as the memory unit, computing the information stored at time $t$, and $h_t$ is the hidden state at time $t$. The parameters $b$ and $\mathcal{\bm W}$ each denote the bias and weight at every level. The activation functions $\sigma_{1}(.)$ and $\sigma_{2}(.)$ are the sigmoid and the hyperbolic tangent function, respectively. The notation $\odot$ represents the element-wise multiplication. During model training, $h_t$ gradually converges towards the model's prediction, $\bm X_{t+1}$. More detailed information about the model's architecture and parameters can be found in~\cite{mdjakirtgcn}.}

\begin{figure*}[!]
    \centering
    \includegraphics[width=2\columnwidth]{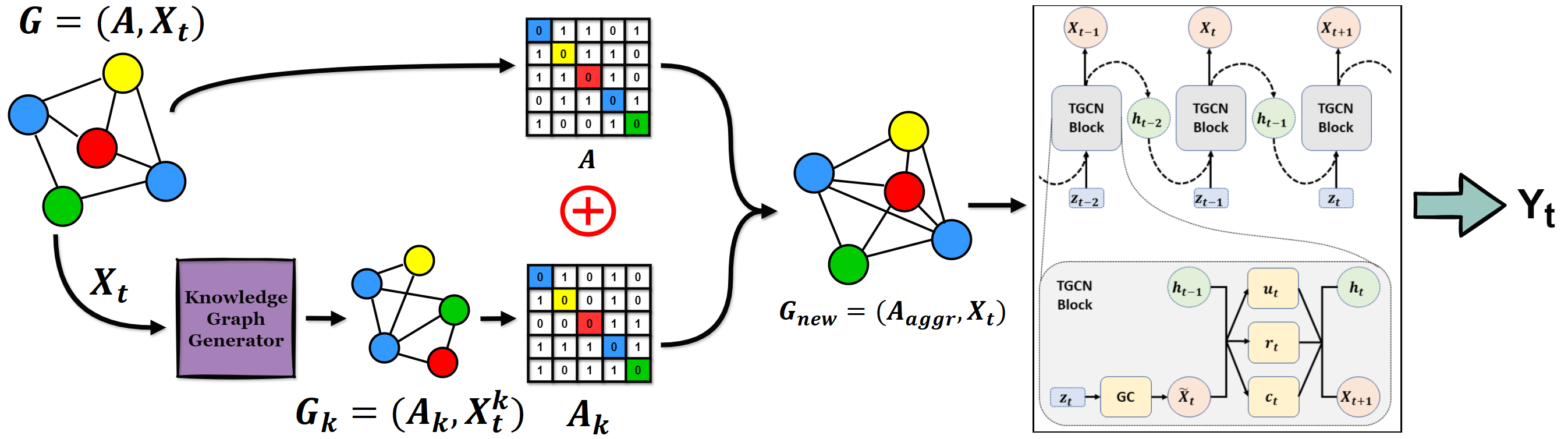}
    \caption{\centering\small{The overview of the Knowledge Graph Infused Model~(KGIM).}}
    \label{scen1}
    \vspace{-0.3cm}
\end{figure*}

\section{Resilient TGCN Framework}
The experiments indicate that the TGCN framework is sensitive to the system's topology, and inaccuracies in topology information can lead to performance degradation in SE. These findings are illustrated under both the original topology and inaccurate topologies with missing or extra links in Figures 3-5 of Section \ref{results}, where they are discussed in detail. In this section, the goal is to revise the reviewed TGCN framework to develop a resilient TGCN for SE that can withstand topology inaccuracies. Specifically, two variations of the model are introduced: Knowledge Graph Infused Model~(KGIM) and Parallel Knowledge Graph Infused Model~(PKGIM). Both models incorporate a knowledge graph, generated from the measurement data, to supplement the potentially inaccurate system topology used in the model. As such, next, the knowledge graph is introduced, followed by two subsections that discuss the KGIM and PKGIM models.

\subsection{Knowledge Graph}\label{kgraph}
Although the true system topology may not be available, the relationships and interactions among measurements can be learned directly from the data. The goal here is to design a {\it knowledge graph} based on system measurements to enhance spatial information processing and learning in the GNN with possibly inaccurate graph input. The knowledge graph is denoted by $\mathcal G_{k}=(\mathcal{V}_k,\mathcal{E}_k)$ with an adjacency matrix ${\bm A_{k}}$. Here, in designing the knowledge graph, it is assumed that the focus is to learn the connections among the known nodes; as such $\mathcal{V}_k = \mathcal{V}$. The learned links between nodes can capture various relationships, such as similarity or correlation among node attributes. Several approaches can be used to define the knowledge graph. Three of these methods are discussed below. 

\subsubsection{Cosine Similarity-based Knowledge Graph}\label{cosine}
A knowledge graph can be defined based on the similarity among the measurement attributes at the nodes of the system. Here, cosine similarity between nodes $i$ and $j$ attributes i.e., phase angles, defined as $s_{ji}=\frac{\mathbf{v_{i}^{T}}\mathbf{v_{j}}}{\|\mathbf{v_i}\|.\|\mathbf{v_j}\|}$, can be exploited to build the knowledge graph, \textcolor{black}{where $\mathbf{v_i}$ represents the attributes of node $i$.} However, using the $s_{ij}$ values alone to define connections in the knowledge graph results in a dense graph with numerous connections of varying similarity weights. To address this, a threshold is necessary to filter out connections with lower similarity values. Specifically, the mean similarity value over all links, denoted as $\bar{s}$, is used to set the threshold, defined as $T = \alpha \bar{s}$. Various $\alpha$ values ranging from 0.75 to 2.75 were tested to fine-tune the knowledge graph, ensuring that the three different approaches of designing knowledge graph produced a similar number of links. The value $\alpha = 2.17$ was found to be effective in achieving this balance.

\subsubsection{Correlation-based Knowledge Graph}\label{correlation}
The knowledge graph can also be defined based on the correlation among the measurements at the nodes of the system. Specifically, the Pearson correlation, defined as $c_{ji}=\frac{\mathbf{(v_{i} - \bar v_{i})^{T}(v_{j} - \bar v_{j})}}{\mathbf{\sqrt{(v_{i} - \bar v_{i})^{T}(v_{i} - \bar v_{i})}\sqrt{(v_{j} - \bar v_{j})^{T}(v_{j} - \bar v_{j})}}}$, can be used to define the connections and their weights between the nodes of the system. Here, $\mathbf{v_i}$ denotes the attributes of node $i$, specifically the phase angle values. Similar to cosine similarity, a threshold is required to filter out links with low correlations in the knowledge graph and to ensure its size remains comparable to the knowledge graphs generated using other techniques. In this work, a threshold of $0.9985$ has been applied for this purpose, resulting in an adjacency matrix with a similar number of edges as in the other two cases.


\subsubsection{Graph Attention Networks~(GAT) Generated Knowledge Graph}\label{gat}
In this approach, the structure of the knowledge graph is learned using a GAT model. Following \cite{gat}, the GAT architecture used here can be defined as:

\begin{equation} \label{gatAttention}
E_{im} = \sigma_{1}(a^{T}(\mathcal{W}O_{i}||\mathcal{W}O_{m}))
\end{equation}
\begin{equation} \label{gatNormalization}
\mu_{im} = \sigma_{2}(E_{im}) = \frac{exp(E_{im})}{\sum_{k} exp(E_{ik})}
\end{equation}
\begin{equation} \label{gatAggregation}
O'_{i} = \sigma_{3}(\sum_{m} \mu_{im}\mathcal{W}O_{m})
\end{equation}

Equation (\ref{gatAttention}) describes the self-attention mechanism used in \cite{gat} to calculate the attention coefficient $E_{im}$. \textcolor{black}{To achieve this, a linear transformation is first applied to the embeddings of node $i$ and its neighbor $m$, which are then concatenated (denoted by $||$), resulting in the weighted combination of the embeddings between node $i$ and its neighbor $m$.}
In this equation, $a$ is a learnable weight vector and $\sigma_{1}(.)$ is the LeakyReLU activation function. The attention coefficients are then normalized using the SoftMax function, $\sigma_{2}(.)$, in Equation (\ref{gatNormalization}) to ensure they sum to 1. Equation (\ref{gatAggregation}) computes each node's output features as a weighted sum of its neighbors' features, using $\mu_{im}$ as weights. \textcolor{black}{$\sigma_{3}(.)$ refers to a non-linear activation function, which in this case is set to the ELU function.} 
By assigning different weights to neighbors, GAT becomes a more flexible and expressive model, providing interpretability through the attention coefficients that indicate each neighbor's importance. GAT addresses the issue of overly dense knowledge graphs generated using the previous two techniques.

In the analyses presented in this paper, the number of links in the knowledge graphs generated using cosine similarity-based methods, correlation-based methods, and the GAT model for the IEEE 118 test case are 227, 226, and 230, respectively (after applying the thresholds). It is also assumed that all links in the knowledge graphs carry equal weight, which is uniformly set to one. In other words, the model does not utilize the weights generated by these approaches for the links. \textcolor{black}{Exploring the use of weighted knowledge links for SE in smart grids can be a direction for future research.} 

\subsection{Knowledge Graph Infused Model~(KGIM) for SE}\label{res-scen1}
The KGIM variation of the TGCN model for SE includes an additional module for generating the knowledge graph using one of the previously mentioned approaches. This knowledge graph is then integrated with the potentially inaccurate initial topology. Specifically, to incorporate the knowledge graph into the initial graph, the elements of the adjacency matrix are updated using a bitwise binary OR operation: $a^{\text{Aggr}}_{ij} = a_{ij} \vee a^{k}_{ij}$ for all $i$ and $j$ in $\mathcal{V}$, where $\vee$ denotes the binary Or operation. Figure~\ref{scen1} illustrates the schematic of the KGIM along with the TGCN components.

\begin{figure*}[!]
    \centering
    \includegraphics[width=2\columnwidth]{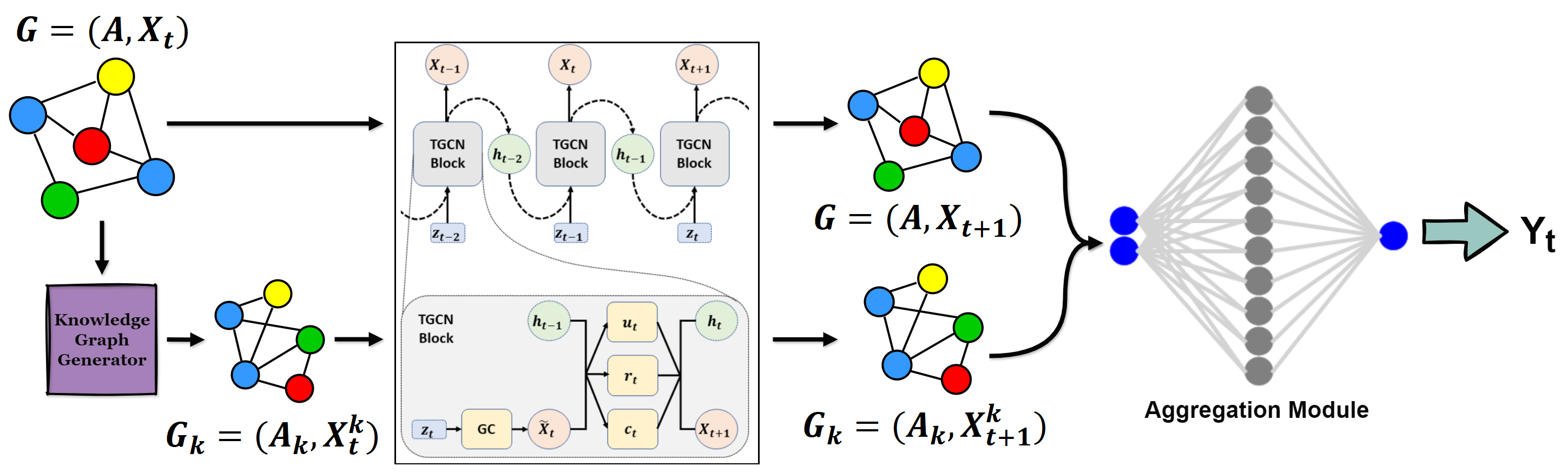}
    \caption{\centering\small{The overview of the Parallel Knowledge Graph Infused Model~(PKGIM).}}
    \label{scen2}
    \vspace{-0.3cm}
\end{figure*}

\subsection{Parallel Knowledge Graph Infused Model~(PKGIM) for SE}\label{res-scen2}
The KGIM model, introduced in the previous subsection, integrates the knowledge graph directly with the initial input topology enabling temporal and spatial learning on the combined topology. In the next model, named PKGIM, the learning process is performed separately and in parallel on the initial topology and the knowledge graph and then the resulting node embeddings are fused into a consolidated state estimate. As the initial input topology is assumed to be close to the true system topology (with inaccuracies affecting only a small number of links), this model allows leveraging the structural information of the initial topology for SE before integrating the knowledge graph; thereby utilizing the graph's structural details more effectively. 
Figure~\ref{scen2} provides an overview of the PKGIM model. The aggregation module plays a crucial role in supporting the TGCN against potential topology inaccuracies by combining the trained node embeddings from both the original graph and the knowledge graph. 
In the following, three approaches for designing aggregation module are discussed. The goal of these designs is to utilize and merge the embeddings from the two graphs through feature learning.

\subsubsection{Single-Layer Perceptron~(SLP) Aggregation}
In the first design for the aggregation module, a simple SLP is utilized as described below.
\begin{equation}\label{nn}
    {\bm Y_{t}}= {\sigma (\mathcal{\bm W}_{3}((\mathcal{\bm W}_{1}\bm \tilde{X}_{1,t})||(\mathcal{\bm W}_{2} \bm \tilde{X}_{2,t})))}
\end{equation}

As shown in equation (\ref{nn}), initially two linear transformations are applied separately to the trained node embeddings from each channel of PKGIM, denoted as $\bm \tilde{X}_{1,t}$ and $\bm \tilde{X}_{2,t}$. These are then concatenated to produce a new embedding with normalized values. The concatenated embedding matrix is then passed through another linear transformation layer, where it is multiplied by $\mathcal{\bm W}_{3}$, followed by a ReLU activation function, $\sigma(.)$. This process results in a weighted output ${\bm Y_{t}}$, which consists of aggregated node embeddings bounded by the range determined by the ReLU function.

\subsubsection{Multi-Layer Perceptron~(MLP) Aggregation} \textcolor{black}{Another aggregation method used in the TGCN framework involves applying an MLP as discussed below.}
\begin{equation}\label{mlp}
 {\bm Y_{t}}={\mathcal{\bm W}_{2}(\sigma (\mathcal{\bm W}_{1}(\bm \tilde{X}_{1,t}||\bm \tilde{X}_{2,t})))}
\end{equation}

A significant difference between SLP and MLP aggregation methods is how the initial linear transformation layer is implemented in the architecture. Here, $\bm \tilde{X}_{1,t}$ and $\bm \tilde{X}_{2,t}$ denote the trained node embeddings of the original and knowledge graphs, respectively, while $\bm Y$ represents the aggregated embeddings. We utilize the ReLU activation function, denoted by $\sigma(.)$, and $\mathcal{\bm W}_{1}$ and $\mathcal{\bm W}_{2}$ referring to the weight matrices used for the linear transformation of the graph features at each layer. Initially, this method concatenates $\bm \tilde{X}_{1,t}$ and $\bm \tilde{X}_{2,t}$ of both the original and knowledge graphs. The concatenated embeddings then undergo a linear transformation, multiplied by $\mathcal{\bm W}_{1}$, followed by the $\sigma(.)$ activation function. 
The final linear transformation layer assigns weighted values $\mathcal{\bm W}_{2}$ to the new embedding matrix from the previous layer, highlighting the importance of specific nodes relative to others. 
This architecture allows the MLP to process the input without requiring explicit normalization at the output. 

\subsubsection{Attention Aggregation} The final data-driven aggregation method implemented in our SE model is ``Attention Aggregation". This method merges node embeddings using learned weights derived from the embeddings, refined through multiple iterations. The process in this model is formulated as:
\[{\bm W}= {\sigma(\mathcal{\bm W}(\bm \tilde{X}_{1,t}||\bm \tilde{X}_{2,t}))} \]
\begin{equation} \label{attention}
  {\bm Y_{t}}= {\sum \bm W^{T} \bm \tilde{X}_{cat}}
\end{equation}

\textcolor{black}{In equation (\ref{attention}), the weight matrix $\bm W$ is learned iteratively in the initial layer of the attention aggregation model. First, the trained node embeddings from the original and knowledge graphs, $\bm \tilde{X}_{1,t}$ and $\bm \tilde{X}_{2,t}$, are concatenated. These concatenated embeddings are then linearly transformed by multiplying with $\mathcal{\bm W}$, followed by applying a SoftMax activation function $\sigma(.)$. This process results in a weighted matrix that integrates features from both graphs. In the subsequent layer, the weighted sum of $\bm W$ with the concatenated embeddings $\bm \tilde{X}_{cat}$ produces $\bm Y$, which is referred to as the ``attention aggregated matrix" in this work.}

With the introduction of the new resilient state estimation models and their architecture, the next step is to present and evaluate their performance under different topology inaccuracies.

\section{Performance Evaluation}\label{results}
\textcolor{black}{This section evaluates the performance of the resilient TGCN framework in scenarios involving topology inaccuracies, such as missing or additional edges in different parts of the system's topology.}

\subsection{Data Generation and Data Preparation}
Here, the IEEE 118 bus system is utilized to generate a large dataset of synthesized PMU time-series through simulations with MATPOWER~\cite{matpower}.
Following the methodology in~\cite{mdjakirtgcn}, the dynamics and temporal aspects are incorporated using load profiles from the New York Independent System Operator (NYISO)~\cite{NYISO}, sampled at 30 Hz. The simulated state variables include time-series data for bus phase angles.
The IEEE 118 bus system has 186 transmission lines and 7 dual links connecting the buses, resulting in a total of 179 links when treating each dual link as individual links. The topology inaccuracy scenarios involving missing and extra edges are structured as follows: (1) {\it Incorrect Missing Edge Scenario:} For each node (bus), one connected line is removed at a time to form $\mathcal G'$. The performance of the TGCN framework is averaged over all missing line scenarios for each node, covering the entire system. (2) {\it Incorrect Extra Edge Scenario}: In each scenario, an additional line is introduced between the target bus and a nearby bus within a radius $\mathcal R$. The radius $\mathcal R=500$ is considered based on geographical coordinates, ensuring proximity between the target and new neighboring buses. \textcolor{black}{This results in 406 inaccurate topologies, each differing from the original by the addition of a single extra link in the bus network graph.}
The TGCN's performance with $\mathcal G'$ is averaged over all extra line scenarios at each node.

\subsection{\textcolor{black}{Performance of the Proposed Resilient TGCN}}
\begin{figure}[htb]
    \centering
    \includegraphics[width=\columnwidth]{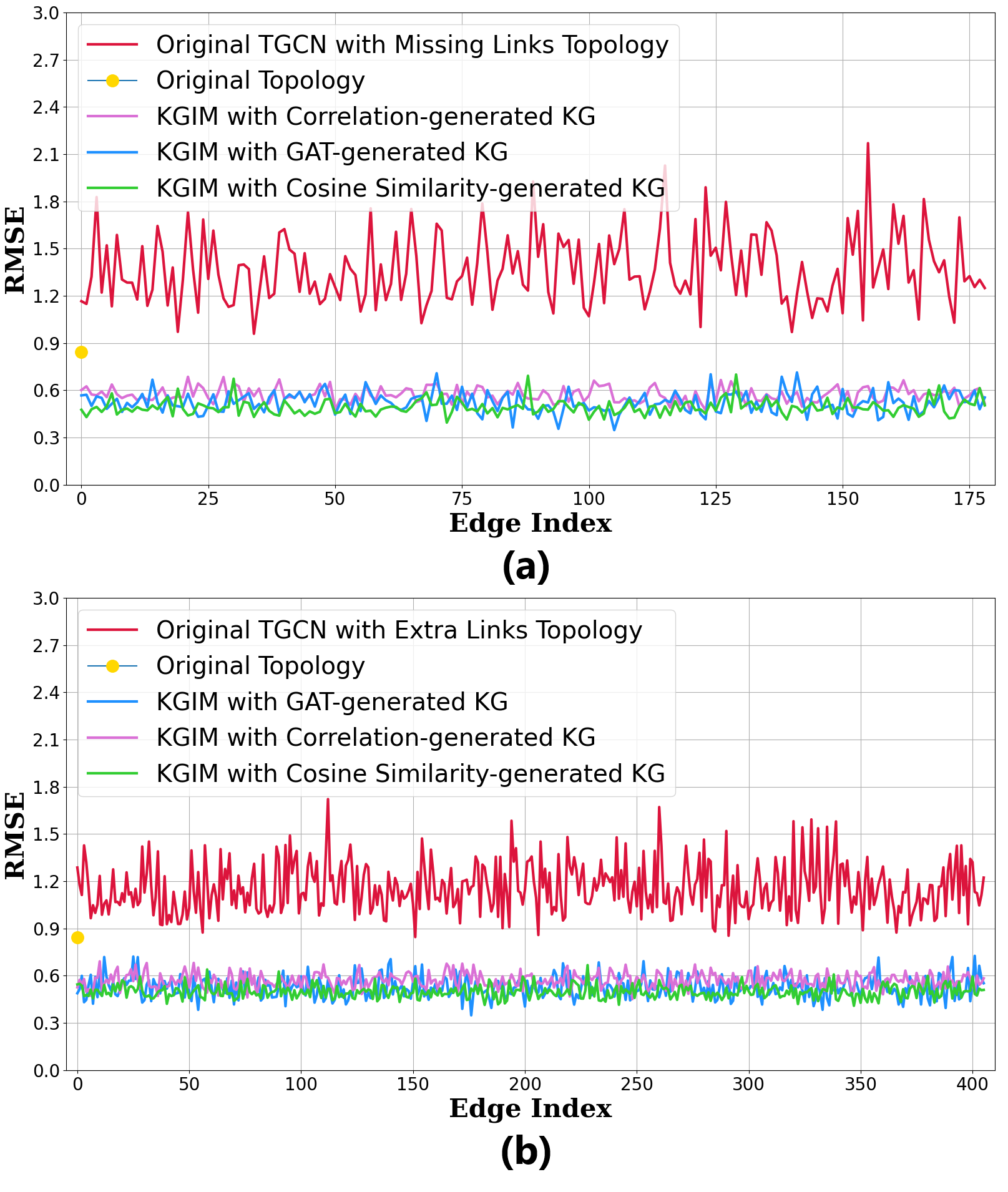}
    \caption{\footnotesize{Performance of KGIM with inaccurate topology scenarios: incorrect missing edges (a) and extra edges (b), using cosine similarity-, correlation- and the GAT model-generated knowledge graphs.}}
    \label{KGIM-results}
    \vspace{-0.1cm}
\end{figure}

\textcolor{black}{In this section, first, the performance of KGIM is evaluated and presented.} Figure~\ref{KGIM-results} shows the KGIM performance under the two aforementioned inaccurate topology scenarios.
\textcolor{black}{The model's performance is evaluated using Root Mean Square Error (RMSE) as the key metric to measure prediction accuracy.}
The red line in the plots represent the original TGCN SE model's performance for different inaccurate topology scenarios and the yellow point indicates the average RMSE of the original TGCN SE model with true topology.

The performance of the KGIM model, using knowledge graphs generated through cosine similarity, correlation analysis, and GAT, is depicted in Figure~\ref{KGIM-results} with green, purple, and blue lines, respectively. As illustrated in the figure, KGIM significantly improves SE performance even in the presence of topology inaccuracies. The results further show that the knowledge graph constructed using cosine similarity has a more favorable impact on the TGCN framework compared to the other two knowledge graphs. This advantage likely stems from the adjacency matrix of the cosine similarity graph, which shares a balanced number of common edges with the original graph (40 edges), compared to the correlation-based graph (118 edges) and the GAT-generated graph (10 edges). This suggests that the cosine similarity graph strikes a better balance between relying on the physical topology and the knowledge graph, enabling the TGCN to perform message-passing on a more robust structure that resembles the original system topology but is different enough to not have major impact from its inaccuracies.

\begin{figure}[htb]
    \centering
    \includegraphics[width=\columnwidth]{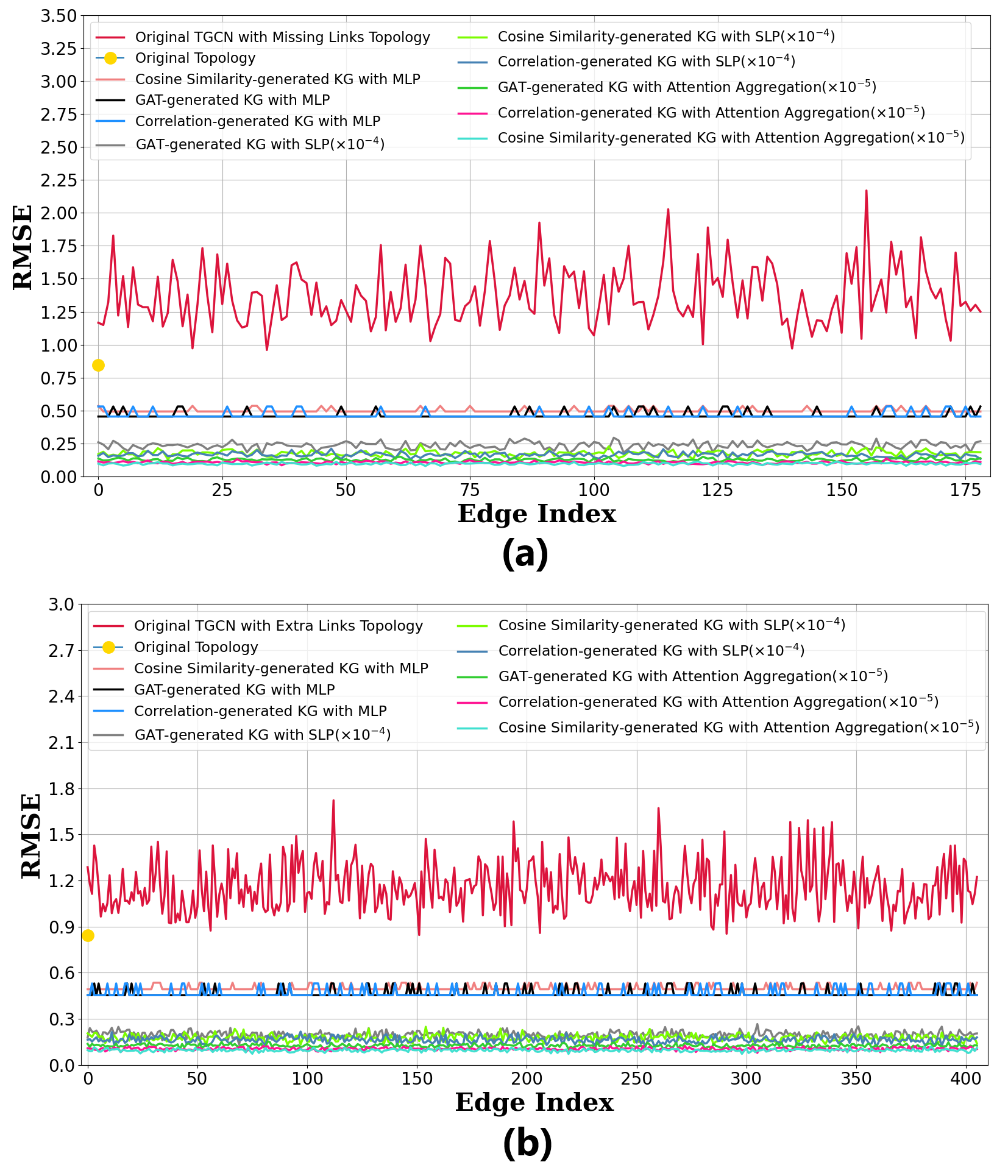}
    \caption{\footnotesize{Performance of PKGIM with inaccurate topology scenarios: incorrect missing edges (a) and extra edges (b), using cosine similarity-, correlation- and the GAT model-generated knowledge graphs.}}
    \label{PKGIM-results}
    \vspace{-0.1cm}
\end{figure}

\begin{figure}[htb]
    \centering
    \includegraphics[width=\columnwidth]{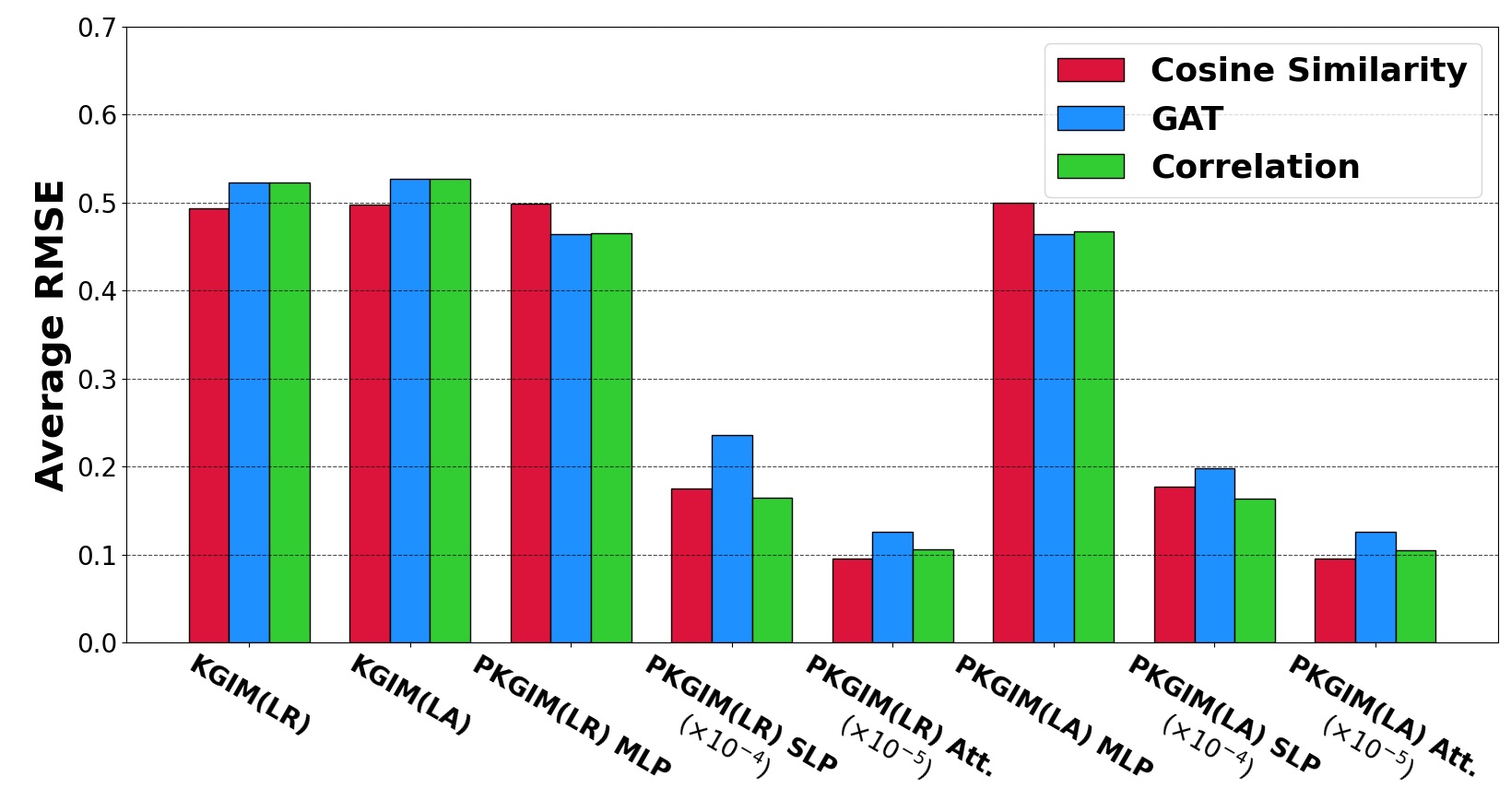}
    \caption{\footnotesize{The overall performance of KGIM and PKGIM with knowledge graphs generate by the cosine similarity method vs. the correlation analysis vs. the GAT model under incorrect missing edges (LR) and extra edges (LA) inaccurate topology scenarios.}}
    \label{overall}
\end{figure}

\textcolor{black}{Unlike KGIM, which focuses on improving the existing TGCN model's resilience to topology inaccuracies by integrating the knowledge graph with the input graph structure, PKGIM operates as a parallel two-channel model. Its goal is to create a robust TGCN architecture by leveraging trained time-series node embeddings from both the original input graph and the knowledge graph, and aggregating them using data-driven methods. As shown in Figure 4, the results from PKGIM, utilizing knowledge graphs generated from the mentioned methods in Section \ref{kgraph}, demonstrate significant improvements in SE performance, with RMSEs reaching magnitudes as low as $10^{-5}$, surpassing KGIM.
\textcolor{black}{Additionally, in Figures 3-4, the original TGCN performance with the true topology, represented by the yellow marker, serves as a baseline for state estimation performance. In both figures, using only the true topology in TGCN exhibits higher RMSE values, indicating that the TGCN captures the underlying relationships between nodes in the original grid with reasonable accuracy for effective state estimation. In contrast, the KGIM and PKGIM, which utilize knowledge graphs, demonstrate significantly better SE performance across all their variants. Their architecture allow them to incorporate additional information about the spatial and structural properties of the grid, which the original TGCN model relying solely on the true topology fails to leverage. This enhanced graph structure enables the KGIM and PKGIM to more effectively capture the correlations between nodes, leading to better error resilience. In simpler terms, the improvement in SE performance seen with the KGIM and PKGIM can be attributed to their ability to construct richer, more informative graph representations with the support of knowledge graphs that capture the underlying node dependencies more accurately than the original TGCN with the true topology.}
Later in this section, the performance of the model with different knowledge graph generation methods across various topology inaccuracies is presented in Figure \ref{overall}.}

\textcolor{black}{The results depicted in Figure~4 highlight the performance of PKGIM with different aggregation modules using the aforementioned knowledge graph generation methods in comparison to the original TGCN under two inaccurate topology scenarios: missing links (Figure 4a) and extra links (Figure 4b). In both cases, the original TGCN with inaccurate topologies (red line) exhibits considerably higher RMSE, demonstrating that disruptions in the network structure, such as missing or additional links, lead to a notable increase in prediction error. This is evident from the elevated RMSE values across various inaccurate topologies. In contrast, the performance of the TGCN using the original true topology (yellow marker) serves as a baseline, consistently showing lower RMSE values below 1, emphasizing the SE performance degradation caused by topology inaccuracies. Integrating the three knowledge graphs via the aggregation module developed in PKGIM show resilience to missing and extra links inaccurate topologies by compensating for the altered network structure and reducing significantly the prediction error. Among the proposed and designed aggregation modules, SLP and attention aggregation methods outperform MLP notably, indicating that deeper architectures does not necessarily provide better robustness in handling topological inaccuracies. Specifically, Figure~4 suggests that the attention aggregation outperforms the other two methods utilized in PKGIM aggregation module and effectively identifies relevant edges and mitigates the impact of noisy or irrelevant connections through to its attention mechanism. Therefore, these results emphasize the effectiveness of integrating knowledge graphs information, generated based on cosine similarity, Pearson correlation, and the GAT model, with the original graph within PKGIM in achieving low RMSE and robust SE performance, even in scenarios where the topological structure is inaccurate.}

To summarize the key findings on the resilience of the KGIM and PKGIM models in the presence of topology inaccuracies, Figure~\ref{overall} illustrates the overall performance of both models using the three knowledge graphs introduced in this study. In the figure, "LR" and "LA" represent "Link Removal" and "Link Addition," respectively, corresponding to scenarios involving incorrect missing links and extra links in inaccurate topologies. 
Results presented in Figure~\ref{overall} indicate that the knowledge graphs generated by cosine similarity and the Pearson correlation method achieve better results compared to those created with the GAT model. Figure~\ref{overall} also confirms that PKGIM is more effective than KGIM in safeguarding the TGCN model against inaccurate topologies. However, within PKGIM, the SLP and attention aggregation methods accomplish significantly better SE performance in the presence of inaccurate topologies compared to the MLP aggregation technique.

\textcolor{black}{As outlined above, Figure~\ref{overall} presents a comprehensive comparison of the performance of both KGIM and PKGIM, evaluated with the three knowledge graphs introduced in this study. One notable takeaway from Figure~\ref{overall}, aside from showcasing the overall performance of the proposed models, is the effectiveness of the GAT-generated knowledge graph in making the TGCN model resilient to topology inaccuracies. Despite the GAT model producing a less dense graph, it achieves nearly the same performance as the cosine similarity-based and correlation-based knowledge graphs in each model. This implies that a data-driven method like the GAT model can deliver strong SE performance with TGCN while avoiding the complexity of a dense graph and the need for manually setting and applying thresholds. This can be attributed to the structure of the GAT-generated knowledge graph. The attention mechanism in the GAT model, defined in Section~\ref{gat}, allows it to construct edges based on the feature relevance of nodes, not necessarily restricted to immediate neighbors. In other words, the GAT model selectively emphasizes important connections, potentially creating a graph that focuses on more relevant relationships than those strictly based on proximity. However, in the case of the cosine similarity-based and correlation-based knowledge graphs, their structures, after applying the threshold, tend to form topologies where each node primarily connects to its immediate neighbors. This results in denser local connections, but the edges are more concentrated around nearby nodes. Consequently, the GAT-based knowledge graph allows the TGCN model to use a graph that has less structural similarity to the original topology, yet still delivers strong performance by focusing on the most important connections.}

\textcolor{black}{As discussed in Section \ref{relatedworks}, the study presented in \cite{TopologyAttackDetection2} is the most similar work to this paper. Consequently, their method of generating a knowledge graph, along with their 4-channel GAT-based SE model has been implemented and evaluated as a benchmark. Using the IEEE 118 test case with phase angles as node features in their framework, an average RMSE of $7.39$ achieved with the true topology. In contrast, the results obtained in this work shows an average RMSE of $10^{-2}$ and $10^{-4}$ in magnitude for the KGIM and PKGIM models, respectively. These findings confirm that the resilient methods proposed and developed in this study have achieved a superior performance in SE compared to the model presented in \cite{TopologyAttackDetection2}.}

\section{Conclusion}
The vulnerability of GCN-based models to noise in underlying graph topology is a critical concern, especially when these models support vital functions like state estimation in power systems. This study aimed to introduce and evaluate various resilience methods to protect GNN-based SE frameworks against different inaccurate topology scenarios, including missing and extra links in the power system graph due to inaccurate data or deliberate adversarial attacks. In this study, two distinct resilient models proposed and designed not only to safeguard but also to enhance the performance of a GNN-based SE frameworks. Central to these models is the integration of an additional graph, known as the knowledge graph, created based on the PMU data. Three methods utilized to generate the knowledge graph: two statistical methods—cosine similarity and Pearson correlation analysis—and a data-driven method utilizing the GAT model. Using the IEEE 118 bus system as the test case, the performance of a temporal GCN SE model analyzed, which incorporates message-passing framework and GRU model for spatial and temporal processing in smart grids. Our findings indicate the effectiveness of these methods in fortifying the TGCN model against various uncertainties in power network. It is quantitatively determined which resilient model outperformed other in terms of mitigating the impact of inaccurate topology. Additionally, the effectiveness and potential of each knowledge graph generated in this study were evaluated and compared within each resilient model, based on the numerical results obtained.

\section*{Acknowledgment}
This material is based upon work supported by the National Science Foundation under Grant No.~2118510.


\end{document}